\documentclass[12pt]{elsart}
\usepackage{amssymb}
\usepackage{amsmath}
\usepackage[dvips]{graphics}
\usepackage{epsfig}
\usepackage{bm}
\expandafter\ifx\csname package@font\endcsname\relax\else
 \expandafter\expandafter
 \expandafter\usepackage
 \expandafter\expandafter
 \expandafter{\csname package@font\endcsname}%
\fi
\begin{document}
\begin{frontmatter}
\title{On the two approaches to the data analysis of the Cassini interplanetary relativity experiment}
\author{Sergei M. Kopeikin}
\address{Department of Physics \& Astronomy, University of
Missouri, Columbia, Missouri 65211}
\ead{kopeikins@missouri.edu}
\date{\today}
\begin{abstract}
We compare two theoretical approaches to the data analysis of the Cassini relativity experiment based on the Doppler tracking and the time delay  technique that were published correspondingly by Kopeikin {\it et al} in Phys. Lett. A {\bf 367}, 276 (2007) and by Bertotti {\it et al} in Class. Quant. Grav. {\bf 25}, 045013 (2008). Bertotti {\it et al} believed that they found a discrepancy with our paper and claimed that our analysis was erroneous.  The present paper elucidates, however, that the discrepancy is illusory and does not exist. The two techniques give the same result making it evident that the numerical value of the PPN parameter $\gamma$ measured in the Cassini experiment is indeed affected by the orbital motion of the Sun around the barycenter of the solar system.
\end{abstract}

\maketitle\end{frontmatter}

\newpage
In 2002 a measurement of the effect of solar gravity upon the phase of
coherent microwave beams passing near the Sun has been carried out with the Cassini
mission, allowing a very accurate measurement of the PPN parameter $\gamma$ \cite{b0,a04}. The data have
been analyzed with NASA's Orbit Determination Program (ODP).
Relativistic ranging time delay, incorporated to the  NASA ODP code, was
originally calculated by Moyer \cite{moy} under assumption that the
gravitating body that deflects light, does not move. Regarding the Sun, it
means that the ODP code derives the ranging delay in the heliocentric frame.
Let us introduce the heliocentric coordinates $X^\alpha=(X^0,X^i)=(cT,{\bm X})$, and use
notation $x^\alpha=(x^0, x^i)=(ct,{\bm x})$ for the barycentric coordinates of the solar
system, which origin is at the center of mass of the solar system. The Sun
moves with respect to the barycentric frame with velocity ${\bm
v}_\odot=d{\bm x}_\odot/dt$ amounting to $\sim 15$ m/s due to the cumulative gravitational attraction of Jupiter, Saturn, and other planets \cite{har}. We have discovered \cite{kpsv} that though this velocity
looks small, it affects the measured value of the PPN parameter $\gamma$ and can not be neglected in the data analysis of such high-precision relativity
experiment as Cassini \cite{b0,a04}.

A legitimate question arises whether the ODP code accounts for the solar
motion or not. We analyzed this question in \cite{kpsv} by doing calculations of the Doppler shift caused by the gravitational field of the moving Sun. We came to the conclusion that the ODP code does take into account the motion of the Sun and that this motion affects the Cassini ranging data. The original papers on Cassini experiment \cite{b0,a04} did not analyze the impact of the solar motion on the results of the experiment. Hence, Bertotti {\it et al} \cite{bai} have decided to repeat our analysis \cite{kpsv} by making direct use of a different approach based on the light-time equation that is the Shapiro ranging time delay. Unfortunately, insufficiently elaborated comparison of the two different mathematical techniques did not allow Bertotti {\it et al} \cite{bai} to reproduce our results. The goal of the present letter is to show that the result of Bertotti {\it et al} paper \cite{bai} exactly coincides with that obtained earlier in our paper \cite{kpsv}.

The ranging time delay in the heliocentric coordinates with the Sun located
at the origin of this frame is well-known \cite{willbook}. After
making use of the heliocentric coordinates it reads
\begin{eqnarray}
\label{tdhf}
T_2-T_1&=&\frac{1}{c}|{\bm X}_2-{\bm X}_1|+\Delta T\;,\\\label{ok}
\Delta
T&=&(1+\gamma)\frac{GM_\odot}{c^3}\ln\left[\frac{R_2+R_1+R_{12}}{R_2+R_1-R_{12}}\right]\;,
\end{eqnarray}
where $\gamma$ is a parameter of the PPN formalism \cite{willbook}, ${\bm X}_2$ and ${\bm X}_1$ are the heliocentric coordinates of
observer on the Earth and emitter (Cassini spacecraft) respectively, distance of the emitter from the Sun is
$R_2=|{\bm X}_2|$, distance of the observer from the Sun is $R_1={\bm X}_1$,
and $R_{12}=|{\bm X}_2-{\bm X}_1|$ is the null-cone heliocentric distance between
the emitter and observer. This equation coincides exactly (after reconciling
our and Moyer's notations for distances) with the ODP time-delay equation
(8-38) given in section 8 of the ODP manual on page 8-19 \cite{moy}.

Moyer
\cite{moy} had transformed the argument of the logarithm in the heliocentric
ranging delay (\ref{ok}) to the barycentric frame by making use of
substitutions
\begin{equation}
\label{jk}
{\bm X}_2\Rightarrow {\bm r}_2={\bm x}_2-{\bm x}_\odot(t_2)\qquad\;,\qquad {\bm X}_1\Rightarrow {\bm r}_2={\bm
x}_1-{\bm x}_\odot(t_1)\;,
\end{equation}
where ${\bm x}_2={\bm x}(t_2)$, ${\bm x}_1={\bm x}(t_1)$ are the barycentric coordinates of the observer and the emitter taken at time of observation, $t_2$, and emission, $t_1$, respectively.
The ODP manual \cite{moy} does not provide any evidence that these
substitutions in the ranging time delay (\ref{ok}) are consistent with relativity and do not violate the Lorentz symmetry of the Cassini experiment. Nonetheless, equations (\ref{ok}), (\ref{jk}) are legitimate transformations from the heliocentric to the
barycentric frame in the sense that they take into account velocity of the
Sun in the {\it argument} of the ranging time delay in the linearized approximation. This is because the solar barycentric coordinate ${\bm x}_\odot(t_1)$ at the time of emission $t_1$ is not the same as the solar coordinate ${\bm x}_\odot(t_2)$ at the time of observation $t_2$
\begin{equation}
\label{zom}
{\bm x}_\odot(t_2)={\bm x}_\odot(t_1)+{\bm v}_\odot(t_1)(t_2-t_1)+O\left(|t_2-t_1|^2\right)\;,
\end{equation}
due to the non-zero velocity ${\bm v}_\odot(t_1)$ of the Sun.

However, spatial transformations (\ref{jk}) are not sufficient in order to get all velocity-dependent terms of the first order in the ranging time delay. The reason is that the Newtonian part of the ranging delay (\ref{tdhf}) contains time in its left side, which must be transformed from one frame to another with taking into account the post-Newtonian correction: $T\Rightarrow t+{\bm v}_\odot{\bm x}/c^2$. This was done by Bertotti {\it et al} \cite{bai} who obtained that the ranging time delay in the heliocentric and barycentric
frames must be related by the simple equation
\begin{equation}
\label{bm}
\Delta t=\left(1+\gamma-{\bm k}\cdot{\bm\beta}_\odot\right) \frac{2GM}{c^3}\ln\left[\frac{r_2+r_1+r_{12}}{r_2+r_1-r_{12}}\right]\;,
\end{equation}
where $r_{12}=|{\bm r}_2-{\bm r}_1|$, $r_1=|{\bm r}_1|$, $r_2=|{\bm r}_2|$, ${\bm\beta}_\odot={\bm v}_\odot/c$, and ${\bm k}$ is a unit vector along the light ray from the emitter to the observer. Formula (\ref{bm}) was derived previously in our work \cite{ks} (see also \cite{kl} for the case $\gamma=1$).

Bertotti et al \cite[section 4]{bai} noticed that our paper \cite{kpsv} neglected the post-Newtonian correction in the transformation of the time coordinate: $T\Rightarrow t+{\bm v}_\odot{\bm x}/c^2$. Hence, \cite{bai} believed that \cite{kpsv} missed the velocity-dependent term in front of the logarithmic function in equation (\ref{bm}). However, our paper \cite{kpsv} dealt with the gravitational Doppler shift of the Cassini radio frequency, that is with the time derivative of the original heliocentric equation (\ref{tdhf}). Transformation of the gravitational Doppler shift does not require to transform time in order to get all linear velocity-dependent corrections, because the heliocentric equation for the Doppler shift is already proportional to velocities of observer and emitter. Hence, the only transformation, which remains to complete is the transformation of the velocities, which can be done after making use of transformations (\ref{jk}) of spatial coordinates differentiated with respect to time (see \cite[equations 12-15]{kpsv}). Our equation for the gravitational Doppler shift, $z_{gr}$,  written down in the barycentric coordinates is \cite{kpsv,ks}
\begin{equation}
\label{qmt}
z_{gr}=-\left(\frac{r_1}{r_{12}}{\bm\beta}_2+\frac{r_2}{r_{12}}{\bm\beta}_1\right)\cdot{\bm\alpha}_B-
\left(\bar\gamma\frac{r_1}{r_{12}}{\bm\beta}_2+\bar\gamma\frac{r_2}{r_{12}}{\bm\beta}_1-{\bm\beta}_\odot\right)\cdot{\bm\alpha}_B\;,
\end{equation}
where ${\bm\beta}_1=(1/c)d{\bm x}_1/dt$ - velocity of the emitter, ${\bm\beta}_2=(1/c)d{\bm x}_2/dt$ - velocity of the observer, $\bar\gamma=\gamma-1$, and the dot between two vectors denotes a scalar dot product. Vector
\begin{equation}
\label{alp}
{\bm\alpha}_B=\alpha_\odot\frac{R_\odot}{d^2}{\bm d}\;,
\end{equation}
where $R_\odot$ - radius of the Sun, ${\bm d}$ - radius-vector of the impact parameter of the radio signal with respect to the Sun, $d=|{\bm d}|$, and $\alpha_\odot=8.5\times 10^{-6}$ rad is the solar gravitational deflection of light on its limb.

The first term in the right side of equation (\ref{qmt}) was obtained by Bertotti and Giampieri \cite{bgam} under condition that the Sun does not move. The second term in the right side of equation (\ref{qmt}) would lead to deviation from general relativity in case of $\bar\gamma\not=0$. The solar velocity $\beta_\odot$ also enters this term, has the same magnitude as the terms with $\bar\gamma$ and, hence, correlates with the measurement of $\bar\gamma$. This $\alpha_B\beta_\odot\simeq 3\times 10^{-13}$ correction comes from the time derivative of the argument of the logarithmic function in equation (\ref{bm}) as shown in \cite{kpsv}. Differentiation of the solar-velocity term in front of the logarithm in equation (\ref{bm}) would give corrections of the next order of magnitude to the right side of the Doppler shit equation (\ref{qmt}). We did not forget that term but neglected it due to its smallness. We conclude that the remark Bertotti {\it et al} \cite{bai} on that we missed the velocity term in front of the logarithm in equation (\ref{bm}) is irrelevant for our Doppler shift formula calculations \cite{kpsv}.

The paper by Bertotti {\it et al} \cite{bai} also claims that the velocity-dependent terms appear in the time delay ({\ref{bm}) only in
front of the logarithmic function. This claim is rather naive and was the reason for misinterpretation of the impact of the solar motion on the Cassini data present in paper \cite{bai}. It only seems like that the argument of the logarithm in equation (\ref{bm}) does not contain velocity-dependent terms explicitly. One should keep in mind that the distance $r_{12}$ is not taken on a single hypersurface of constant time but connects two different positions of the Sun, which are not the same in the barycentric coordinates because the Sun moves. The post-Newtonian expansion of distance $r_{12}$ yields
\begin{equation}
\label{p14q}
r_{12}=r-{\bm r}\cdot{\bm\beta}_\odot+O(\beta^2_\odot)\;,
\end{equation}
where $r=|{\bm r}|$ and ${\bm r}={\bm x}_2-{\bm x}_1={\bm k}r$ is a null-cone vector connecting the observer to the emitter along the radio wave path from the Cassini to the observer. The post-Newtonian expansion of the ranging delay (\ref{bm}) is
\begin{equation}
\label{p15}
\Delta t=(1+\gamma-{\bm
k}\cdot{\bm\beta}_\odot)\frac{2GM}{c^3}\ln\left[\frac{r_2+r_1+r-{\bm
r}\cdot{\bm\beta}_\odot}{r_2+r_1-r+{\bm
r}\cdot{\bm\beta}_\odot}\right]+O\left(\frac{2GM\beta^2_\odot}{c^3}\right)\;,
\end{equation}
which explicitly reveals the presence of the velocity-dependent terms in the argument of the ranging time delay. Equation (\ref{p15}) has been derived in our paper \cite{kpsv}, and its logarithmic part is just a partial derivative of the {\it heliocentric} time delay with respect to the PPN parameter $\gamma$ as shown in \cite[equation 25]{kpsv}.

Bertotti et al. \cite{bai} claimed that expression (\ref{p15}) for the
ranging time delay is not used in the ODP code and can not be applied for
theoretical analysis of the Cassini experiment as we did in \cite{kpsv}.
Therefore, Bertotti et al. \cite{bai} have concluded that our numerical
estimates of the gravitational shift of frequency caused by motion of the
Sun with respect to the barycenter of the solar system as given in
\cite{kpsv}, are incorrect. These statements of Bertotti et al. \cite{bai} can not be accepted by any rationally thinking researcher
as the authors of \cite{bai} have trivially overlooked that expression (\ref{p15}) has exactly the same
logarithmic function as in equation (\ref{bm}) with the argument expressed in terms of the
null-cone distance $r$ and velocity of the Sun, ${\bm v}_\odot$, which are related to distance $r_{12}$  via self-consistent mathematical transformation (\ref{p14q}).
The equivalence of equations (\ref{bm}) and (\ref{p15}) means that our numerical estimates and theoretical conclusions given in paper
\cite{kpsv} with regard to the impact of the solar motion on the Cassini measurement of $\gamma$ parameter, are firmly confirmed by the independent study of Bertotti {\it et al} \cite{bai}.

\end{document}